\documentclass[10pt]{iopart}

\usepackage{graphicx}
\usepackage{bm}
\usepackage[usenames, dvipsnames]{color}
\usepackage{multirow}
\usepackage{cite}
\usepackage[colorlinks=true,linkcolor=blue,urlcolor=blue,citecolor=blue]{hyperref}

\begin{document}

\title{Characterization of Majorana-Ising phase transition
in a helical liquid system}
\author{Sudip Kumar Saha$^1$, Dayasindhu Dey$^1$, Monalisa Singh Roy$^1$, Sujit Sarkar$^2$ and Manoranjan Kumar$^1$}
\address{$^1$ S. N. Bose National Centre for Basic Sciences, Block - JD, Sector - III, Salt Lake, Kolkata - 700106, India}
\address{$^2$ Poornaprajna Institute of Scientific Research, 4 Sadashivanagar, Bangalore 560080, India}
\eads{\mailto{sudipksaha@bose.res.in}, \mailto{dayasindhu.dey@gmail.com}, \mailto{singhroy.monalisa@gmail.com}, \mailto{sujit.tifr@gmail.com}, \mailto{manoranjan.kumar@bose.res.in}}

\date{\today}
\begin{abstract}
We map an interacting helical liquid system, coupled to an
external magnetic field and s-wave superconductor, to an XYZ 
spin system, and it undergoes Majorana-Ising transition 
by tuning of parameters. In the Majorana state lowest 
excitation gap decays exponentially with system size, and the 
system has degenerate ground state in the thermodynamic limit. 
On the contrary, the gap opens in the Ising phase even in the 
thermodynamic limit. We also study other criteria to characterize 
the transition, such as edge spin correlation with its neighbor 
$C(r=1)$, local susceptibility $\chi_i$, superconducting 
order parameter  of edge spin $P(r=1)$, and longitudinal 
structure factor $S(k)$. The ground state degeneracy and three 
other criteria lead to the same critical value of parameters 
for Majorana-Ising phase transition in the thermodynamic limit. 
We study, for the first time, the entanglement spectrum of the reduced 
density matrix of the helical liquid system. The 
system shows finite Schmidt gap and non-degeneracy of the 
entanglement spectrum in the Ising limit. The Schmidt gap 
closes in the Majorana state, and all the eigenvalues are 
either doubly or multiply degenerate.
\end{abstract}


\maketitle
\ioptwocol
\section{\label{sec1}Introduction}
The unique properties of the Majorana fermion behaving as its 
own anti-particle and the existence of particle and 
anti-particle in the same system are the subjects of intense 
research interest in various branches of physics, 
especially in quantum many-body systems since its proposal 
by  E. Majorana~\cite{majo1,wil,berni}. The recent studies 
show that the concept and existence of Majorana, a zero energy 
mode, like quasi-particles is very common in many branches of 
physics. Interestingly, it appears as an emergent particle in 
various condensed matter systems, such as Bogoliubov 
quasi-particle in a one-dimensional 
superconductor~\cite{ivanov, kraus, wimmer, sato}, semiconductor 
quantum wire~\cite{stanescu2013, mourik, deng, rokin, das, finc, jafari}, proximity 
induced topological 
superconductor~\cite{kitaev, fu1, fu2, ali1, ali, sau, potter, ali2, fukui, sarkar3}, 
and the cold atoms trapped in one-dimension~\cite{zhang, jiang}. 
Other than the exotic physics of this topological state, the absence 
of decoherence in Majorana fermionic system makes it a prospective candidate 
for application in the non-abelian quantum 
computation~\cite{nayak, kitaev, sau2, ali3}. 

The real world applications of these fermionic systems depend on 
the stability of the Majorana fermions. A single spin polarized 
fermion band system with spin orbit coupling and proximity induced 
superconductor shows Majorana like modes in the presence of weak 
fermionic interaction. However, it was shown that the interaction 
weakens the stability of the Majorana fermion~\cite{suhas}. 
Potter and Lee~\cite{potter} showed that the $p+ip$ superconductor 
possesses localized Majorana particles in a rectangular system with 
width less than the coherence length of the superconductor. 
The helical liquid system is another candidate where the Majorana 
like quasi-particles can exist. The helical liquid system generally 
originates because  of the quantum spin Hall effect in a system with 
or without Landau levels. In this system, a coupling of the left 
moving down spin with the right moving up spin at the edge of two 
dimensional quantum hall systems gives rise to a quantized transport 
process. In this phase, the spin and the momentum degrees of freedom are 
coupled together without breaking the time reversal symmetry. 
Various aspects of helical spin liquid is discussed in~\cite{sela,sarkar2,qi,qi2,dspms}.

The field theoretical calculation by Sela et al.~\cite{sela}  shows 
that the Majorana bound state in a helical liquid possesses a higher 
degree of stability. In presence of the interaction, the scattering 
processes between the two constituent fermion bands in the helical 
liquid system stabilizes the Majorana bound state by opening a 
gap~\cite{suhas,ali2,sela,sarkar2}. However, the strong interaction 
may induce decoherence in the Majorana modes. They also considered 
a highly anisotropic spin model with transverse and longitudinal 
fields, and the system shows Majorana to Ising transition 
(MI)~\cite{sela,sarkar2,dspms}. One of our coauthors showed using 
RG calculation that a transition from a phase with Majorana edge 
modes to the Ising phase exists in the helical liquid system for both 
presence and absence of interaction~\cite{sarkar2}. However, a 
systematic and accurate calculation of the phase boundary of the 
MI quantum phase transition is still absent in the literature. 

In this paper, we study the existence of Majorana fermion 
modes in the interacting helical liquid system, and also propose 
different criteria to characterize the MI quantum phase transition. 
We will also analyze the entanglement spectrum (ES) to characterize 
the topological aspects of the Majorana edge modes.

Sela et al.~\cite{sela} introduced a helical fermionic system, which 
can be written in the field theoretical representation as
\begin{eqnarray}
H&=& H_0+\delta H+ H_{fw}+ H_{um}, 
\label{eq:h0}
\end{eqnarray}
where $H_0$ and $\delta H$  terms include the kinetic energy, single 
potential energy, external magnetic field, and proximity induced 
energy  terms. $H_{fw}$ and $H_{um}$ represent the forward scattering 
and the umklapp scattering terms respectively. The $H_0$ and $\delta H$ 
terms can be written as 
\begin{eqnarray}
H_0&=&\int dx \left[\psi_{L \downarrow}^{\dagger} 
(v_F i \partial_x - \mu)\psi_{L \downarrow} \right. \nonumber \\
& & \qquad + \left. \psi_{R \uparrow}^{\dagger} 
(-v_F i \partial_x - \mu)\psi_{R \uparrow} \right],\nonumber \\
\delta H &=& \int dx \left[  B \psi_{L \downarrow}^{\dagger}\psi_{R \uparrow} 
+ \Delta {\psi_{L \downarrow}} {\psi_{R \uparrow}} + h.c.\right],
\label{eq:h0p}
\end{eqnarray}
where $ \Psi $ are field operators, $ v_F $ and $ \mu $ are the 
Fermi-velocity and chemical potential of the helical liquid. The 
system is coupled to the magnetic field $ B $, and proximity 
coupled to $ s $-wave superconducting gap $ \Delta $ which is 
shown by additional terms in the Hamiltonian.
Both the scattering terms are given as
\begin{eqnarray}
\label{eq:scatter}
H_{fw} &=& \int dx \left[ g_2  \psi_{L \downarrow}^{\dagger} \psi_{L \downarrow}
 \psi_{R \uparrow}^{\dagger} \psi_{R \uparrow}  \right. \nonumber \\
& & + \left. \frac{g_4}{2} \left\{ \left( \psi_{L \downarrow}^{\dagger} 
\psi_{L \downarrow}\right)^2 + \left( \psi_{R \uparrow}^{\dagger} 
\psi_{R \uparrow}\right)^2  \right\} \right],  \nonumber \\
H_{um} &=& g_u \int dx \left[ \psi_{L \downarrow}^{\dagger} \partial_x \psi_{L \downarrow}^{\dagger}
\psi_{R \uparrow} \partial_x \psi_{R \uparrow} + h.c.\right]. 
\end{eqnarray}
The conventional analytical expression for the umklapp scattering 
term $H_{um}$ for the half filling~\cite{wu,sela,sarkar2,dspms} 
is written in \eref{eq:scatter}. This analytical expression gives a 
regularized theory using the lattice constant $a$ as an ultraviolet cut-off.

This complete Hamiltonian $ H + \delta H + H_{fw} + H_{um} $ can 
be mapped to a spin-$1/2$ $XYZ$ model 
Hamiltonian~\cite{sela, sarkar2, dspms} and can be written  
as~\cite{sela}  
\begin{equation}
H_i = \sum_{i} J^{\alpha} S_i^{\alpha} S_{i+1}^{\alpha}
- \left[  \mu + B (-1)^i \right]  S_i^z ,
\label{sela:ham}
\end{equation}
where $\alpha\equiv x,y,z$ and $ J^\alpha $  are different 
components of  spin exchange  interaction between neighboring 
spins. $ \mu $ and $ B $ are longitudinal normal and staggered  
magnetization applied externally. Here, we assume $J^x=J+\Delta$ 
and $ J^y=J-\Delta $ where $ \Delta $ is the superconducting gap. 
\Eref{sela:ham} can be rewritten in terms of new variables 
as
\begin{eqnarray}
\label{eq:spin_ham}
H &=& \frac{J}{2} \sum_{i}  \left( S_{i}^{+} S_{i+1}^{-} + S_{i}^{-} S_{i+1}^{+}\right) + 
J^{z} \sum_{i} S_{i}^{z} S_{i+1}^{z}  \nonumber \\
& &+ \frac{\Delta}{2} \sum_{i}   \left( S_{i}^{+} S_{i+1}^{+} +S_{i}^{-} S_{i+1}^{-}\right)  - 
 \sum_{i} \left[  \mu + B(-1)^i \right]  S_{i}^{z} . \nonumber \\
\end{eqnarray} 
To have a better understanding of the Majorana modes, we map 
this spin system to the spinless fermion model using Jordan-Wigner 
transformation~\cite{fradkin}  of \eref{eq:spin_ham}, and 
it can be written in terms of spinless fermion as~\cite{katsura1}
\begin{eqnarray}
\label{eq:fermion_ham}
 H &=& \frac{-J}{2} \sum_{i} \left( c_{i}^{\dagger} c_{i+1} + h.c.\right)  \nonumber \\
& & + J^{z} \sum_{i}  \left( c_{i}^{\dagger}c_i -\frac{1}{2}\right) 
\left( c_{i+1}^{\dagger}c_{i+1} -\frac{1}{2}\right)  \nonumber \\ 
& & + \frac{\Delta}{2} \sum_{i}  \left( c_{i+1}^{\dagger} c_{i}^{\dagger} + h.c.\right)  \nonumber \\ 
& &- \sum_{i} \left[  \mu + B(-1)^i \right] \left( c_{i}^{\dagger}c_i -\frac{1}{2}\right)  .
\end{eqnarray} 

For the sake of completeness, let us try to understand the results 
in the limiting cases. This model is well studied in the limit of 
$B=0$ and $ J^z=0 $, and \eref{eq:fermion_ham} reduces to 
1D Kitaev  model~\cite{kitaev,katsura2} which is given by
\begin{eqnarray}
\label{eq:kitaev_ham}
 H &=& -\frac{J}{2} \sum_{i} \left( c_{i}^{\dagger} c_{i+1} + h.c.\right)  \nonumber \\ 
& &+\frac{\Delta}{2} \sum_{i}  \left( c_{i+1}^{\dagger} c_{i}^{\dagger} + h.c.\right) - 
\mu \sum_{i}  \left( c_{i}^{\dagger}c_i -\frac{1}{2}\right)  .
\end{eqnarray} 

Now let us consider a transformation $ {c_j} = \frac{1}{2} ( a_{2j-1} + i a_{2j} )$ 
and 
$ c_j^{\dagger} = \frac{1}{2} ( a_{2j-1} - i a_{2j} )$. Here, $ a_j^{\dagger} $ 
and $ a_j $ are the creation and annihilation operators of  $j^{\rm{th}}$ Majorana fermion.  
We can write~\eref{eq:kitaev_ham} as
\begin{eqnarray}
\label{eq:reduced_kitaev_ham}
H&=&\frac{i}{2} \sum_{j} [ - \mu a_{2j-1}a_{2j} +\frac{1}{2} \left( J+\Delta\right) a_{2j}a_{2j+1} 
\nonumber \\ 
& & \qquad +\frac{1}{2} \left( -J+\Delta\right)  a_{2j-1} a_{2j+2} ] .
\end{eqnarray}
There are two conditions: first, when $J=\Delta=0$ and $\mu < 0$, 
system shows trivial phase and two Majorana operators at each site 
are paired together to form a ground state with occupation number $0$. 
Secondly, for  $J=\Delta>0 $ and $ \mu=0 $, the Majorana 
operators from two neighboring sites are coupled together leaving two 
unpaired Majorana operators at the two ends, and these two Majorana 
modes are not coupled to the rest of the chain~\cite{kitaev,ali}. 
The fermionic edge state formed with these two end operators has 
occupation 0 or 1 with degenerate ground state, i.e., generating zero 
energy excitation modes. However, the bulk properties of these systems 
can be gapped. 

In most of the papers~\cite{gergs,affleck_dmrg,ali2,haldane_dmrg,satoshi}, 
the Majorana zero modes (MZM) are characterized by exponential decay of 
the lowest excitation gap with system size, and a large expectation value 
of the creation operator of the  fermion $\langle GS|c^{\dagger}|GS \rangle$ 
near the edges of the system. However, in the spin language there is no trivial relation 
between the Majorana mode and spin raising operators. Therefore, in this 
paper our main focus is to  find the accurate phase boundary of the MI 
transition, and for this purpose we focus on the lowest excitation gap 
$ \Gamma $, derivative of longitudinal spin-spin correlations $C(1)$ 
between edge spin and its nearest neighbor spin, and spin density $ \rho_e $ 
of edge sites. In principle, the MZM do not couple with the bulk 
states~\cite{kitaev,ali}; therefore, this correlation of the 
edge spin should decay exponentially. We notice that a local  magnetic 
susceptibility $ \chi_i $  shows a discontinuity near the phase transition. 
Fifth quantity is $ P(r) $ of edge spin, and it is similar to the spin 
quadrupolar/spin-nematic order parameter~\cite{nematic1,nematic2,nematic3} 
or superconducting order parameter of model in~\eref{eq:fermion_ham}. 
The structure factor $S(q)$ gives us information about the phase boundary 
and the bulk state. Based on the above quantities the MI transition 
boundary is calculated in this paper. The bulk properties of the Majorana 
state is rarely discussed in the literature, however we will try to 
discuss those properties in this paper. In later part of this paper, 
the ES of both states are also discussed to understand the topological 
aspects of Majorana modes. 

The Hamiltonian mentioned in~\eref{eq:spin_ham} is solved using the 
Density matrix renormalization group (DMRG)~\cite{white-prl92, white-prb93} 
and the exact diagonalization (ED) method. The DMRG method is a state of 
the art numerical technique to solve the $1D$ interacting system, and 
is based on the systematic truncation of irrelevant degrees of freedom 
in the Hilbert space~\cite{white-prl92, white-prb93}. This numerical 
method is best suited to calculate a few low lying excited states of 
strongly interacting quantum systems accurately. To solve the interacting 
Hamiltonian for ladder and chain with periodic boundary condition, the 
DMRG method is further improved by modifying conventional DMRG 
method~\cite{dd2016a} for zigzag chains~\cite{mk2010}, quasi one 
dimension~\cite{mk2016} and higher dimensions~\cite{mk2012}. The left 
and right block symmetry of DMRG algorithm for a XYZ-model of a 
spin-$1/2$ chain in a staggered magnetic field (in \eref{eq:spin_ham}) 
is broken. Therefore, we use conventional unsymmetrized DMRG 
algorithm~\cite{white-prl92, white-prb93} with open boundary condition. 
In this model, the total $S^z$ is not conserved as $S^z$ does not commute 
with the Hamiltonian in~\eref{eq:spin_ham}. As a result, the 
superblock dimension is large. We keep $m \sim 500$ eigenvectors 
corresponding to the highest eigenvalues of the density matrix to 
maintain the desired accuracy of the results. The truncation error 
of density matrix eigenvalues is less than $10^{-12}$. The energy 
convergence is better than $0.001\%$ after five finite DMRG sweeps.  
We go upto $N=200$ sites for the extrapolation of the transition points. 

  This paper is divided in to three sections. In~\sref{sec2}, 
we discuss our numerical results, and this is divided into eight 
subsections.  Results are discussed and compared with the existing 
literatures in~\sref{sec3}. 
\section{\label{sec2}Numerical results}
In this section, various criteria for the MI transition are discussed. 
We start with a three dimensional phase diagram in $B$, $ \Delta $ 
and $ \mu $  parameter space for given $J^z =0$ and $0.5$.  Thereafter, 
various criteria of the MI such as lowest excitation energy $ \Gamma $, 
edge spin correlation with its nearest neighbor $C(r=1)$, local 
susceptibility at the site nearest neighbor to the edge $\chi_2$, 
superconducting or spin-nematic  order parameter  of edge spin $P(r=1)$,  
and structure factor $S(k)$ are studied. We show that all these 
quantities show extrema at a transition parameter $B_m$ in $\Delta-B$ 
parameter space. However, all these extrema $B_m$ are extrapolated to 
the same point in the thermodynamic limit, and this extrapolation is 
done in the~\sref{Sec:H}. The ES is analyzed in~\sref{Sec:G} 
to show the distinction between topological and the Ising phase. 

\subsection{\label{Sec:A}Phase diagram}
%
\begin{figure}
\begin{center}
\includegraphics[width=3.3 in]{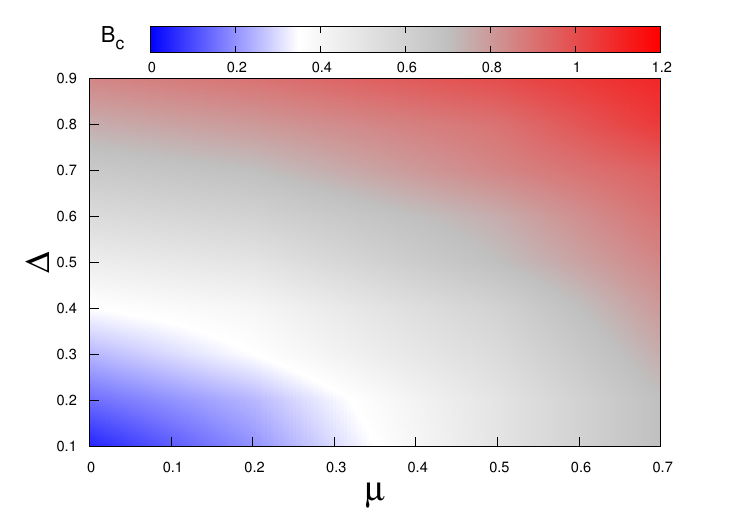}
\caption{MI phase boundary for a helical liquid mentioned 
in~\eref{eq:spin_ham} in the  parameter space of $\Delta $ and 
$ \mu $ for $J^z = 0$. The color gradient represents the 
critical magnetic field $B_c $ in the phase boundary.}
\label{fig1}
\end{center}
\end{figure}
%
A phase diagram of the model Hamiltonian in~\eref{eq:spin_ham}, 
is shown as a color gradient plot in~\fref{fig1}, where the 
color gradient represents the critical value of $B_c$, and $X$- 
and $Y$-axis are the $ \mu $ and the $ \Delta $ of the MI transition 
points for a system size $N=100$ for $J^z=0$. The phase transition 
in the $ \mu-\Delta $ parameter space shows that a finite $ \Delta_c $ 
is required to generate the Majorana modes in a finite system. The 
$\mu$ favors the longitudinal degrees of freedom and tries to induce 
the ferromagnetic order, although $B$ tries to align the nearest 
spins in opposite directions to induce the Antiferromagnetic N\'eel 
phase. In fact the $B$ and the $\mu$ both favor the Ising order, 
whereas the $ \Delta $ breaks the parity symmetry and induces the 
degeneracy in the system. It also induces the formation of 
Cooper-pairs or magnon-pairs like excitations at the two neighboring 
sites for the model Hamiltonian given by~\eref{eq:fermion_ham} 
or~\eref{eq:spin_ham} respectively. As shown in~\fref{fig1}, 
the transition value $B_c$  increases with increasing 
$ \mu_c $ for a fixed value of $ \Delta_c $, 
and it increases with increasing $ \Delta_c $ for a given $ \mu_c $, 
and the trends are similar for both $J^z=0$ and $0.5$. We notice 
that the Majorana state occurs for $(B-\mu)^2 < \Delta^2$.   

\subsection{\label{Sec:B}Excitation gap $ \Gamma $}
\begin{figure}
\begin{center}
\includegraphics[width=3.3 in]{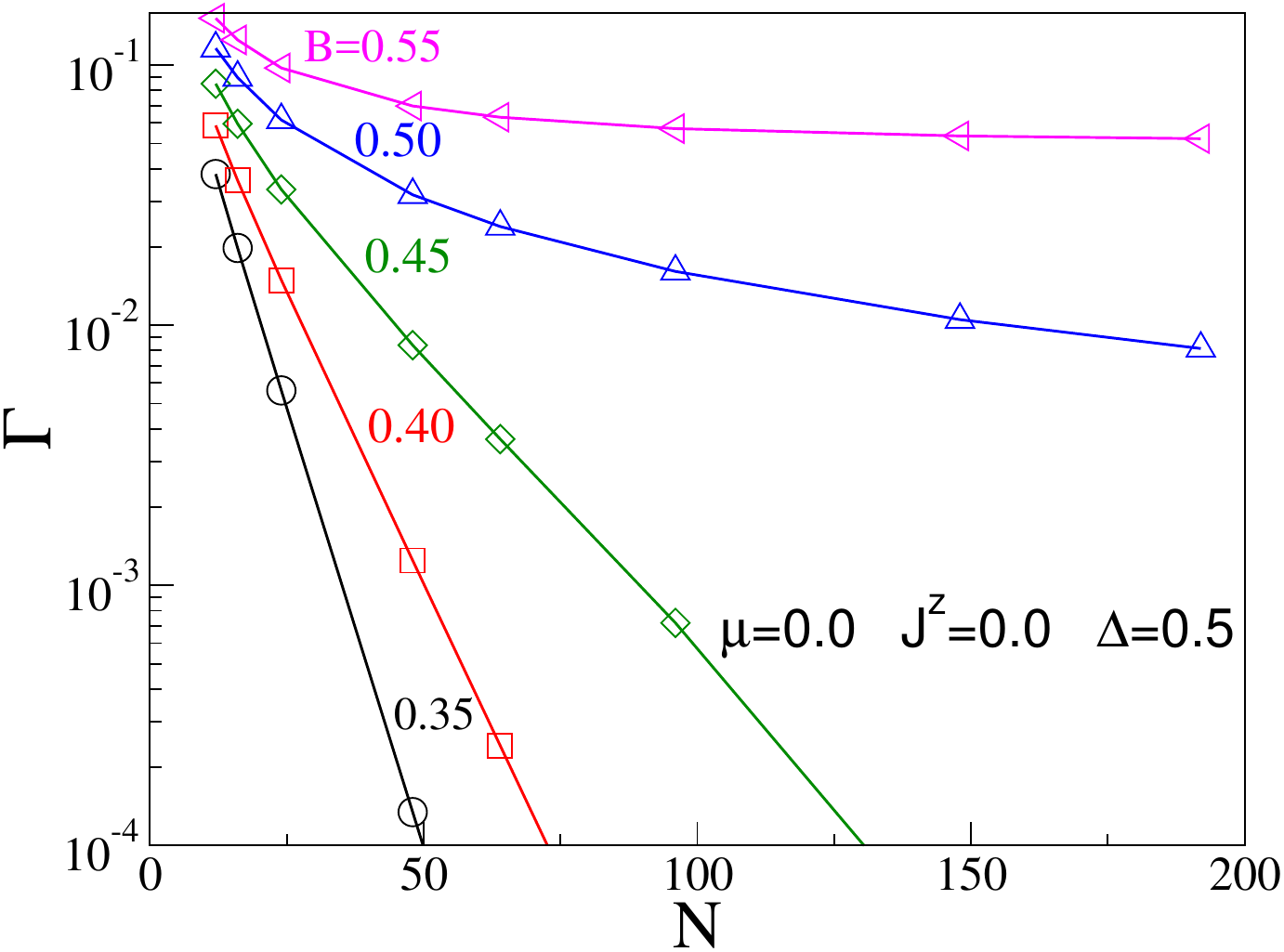}
\caption{Lowest excitation gap $\Gamma$ (in~\eref{eq:gap}) 
vs. the system size $N$ for $\mu = J^z = 0$ and $\Delta=0.5$
with different values of $B$ chosen around the $B_c = 0.48$ 
(see~\fref{fig1}).}
\label{fig2}
\end{center}
\end{figure}

To characterize the Majorana modes in one dimension for the helical 
system, the lowest excitation gap $ \Gamma $ is defined as
\begin{equation}
\Gamma=E_1(\Delta,\mu,B)-E_0(\Delta,\mu,B),
\label{eq:gap}
\end{equation}
where $ E_0 $ and $E_1$ are ground state and lowest excited state 
of the Hamiltonian in~\eref{eq:spin_ham}. The  $\Gamma-N$ is 
plotted in log-linear scale in~\fref{fig2}. 
The $\Gamma-N$ plot for $\mu = 0$, $ \Delta = 0.5 $, $J^z = 0.0$, 
and for five values of $B$. For $B = 0.35$, $0.4$ and $0.45$,  
$ \Gamma $ shows the exponential decay (Majorana regime), whereas 
$ \Gamma $ goes linearly with $1/N$ for $B=0.5$ and $0.55$ 
(Ising phase) as shown in the~\fref{fig2}. The phase boundary 
of MI transition is evaluated based on change in $\Gamma-N$ relation 
from exponential to the power law.  We also notice that the 
contribution to the excitation gap $ \Gamma $ is uniformly 
distributed in the Ising phase, whereas, in the Majorana state, 
the major contribution comes from the edge as shown in Fig.~5 of~\cite{dspms}. 
%
\subsection{\label{Sec:C}Correlation function $C(r=1)$ from the edge}
\begin{figure}
\begin{center}
\includegraphics[width=3.3 in]{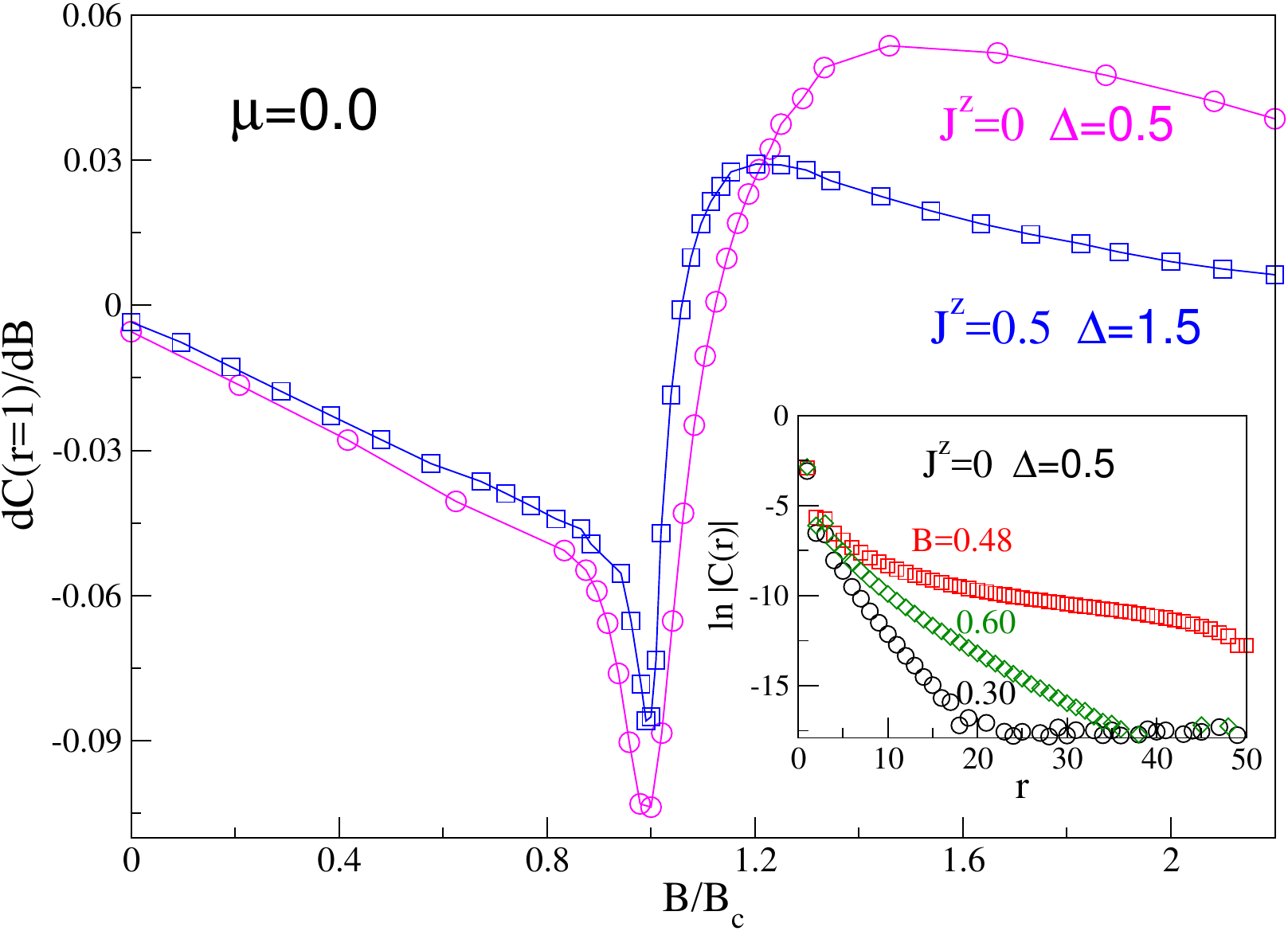}
\caption{The derivative of longitudinal spin-spin correlation 
between the spins corresponding to the edge bond $dC(r=1)/dB$ 
with the staggered magnetic field $B$ both for ($\Delta = 0.5$, 
$\mu  = J^z = 0$) and ($\Delta = 1.5$, $J^z = 0.5$, $ \mu =0 $). 
The inset shows the distance dependence of $\ln |C(r)|$ for 
$B = 0.3$, $0.48$ and $0.6$ and for $\Delta = 0.5$, $\mu  = J^z = 0$.}
\label{fig3}
\end{center}
\end{figure}

The longitudinal spin-spin correlation fluctuation $C(r)$ at 
a distance $r$ from reference point $i$ is  defined as 
\begin{equation}
\label{spin_spin_corl}
C(r)=\left\langle S^z_iS^z_{i+r}\right\rangle - 
\left\langle S^z_i\right\rangle \left\langle  S^z_{i+r}\right\rangle,
\end{equation}
where $ \left\langle S^z_i\right\rangle$ and 
$\left\langle S^z_{i+r}\right\rangle$ are spin densities at 
the reference site $i$ and other site $i + r$. In~\fref{fig3}, 
the edge spin site $i = 1$ is considered as the reference spin. 
The distance dependence of $C(r)$ for $\mu = 0$, $J^z=0$ and 
$\Delta=0.5$ is shown in inset of~\fref{fig3}. It decreases 
exponentially for $r \geq 2$, and effectively, only last two sites 
are correlated. Therefore, $C(r=1)$ between nearest neighbors is 
important. $C(r=1)$ first increases with $B$  in the Majorana state 
and decreases afterwards in the Ising phase. The $dC(r=1)/dB$ is 
plotted as a function of $B/B_c$ in the main~\fref{fig3} for 
$\left( J^z=0\right. $, $\left.  \Delta=0.5\right)  $, and 
$\left( J^z=0.5\right.$, $ \left. \Delta=1.5 \right) $ and $ \mu=0 $. 
This minimum for the given value of parameters is also consistent 
with the MI transition point calculated from energy degeneracy.
%
%
\subsection{\label{Sec:D}Local magnetic  susceptibility $\chi_i$}
\begin{figure}
\begin{center}
\includegraphics[width=3.3 in]{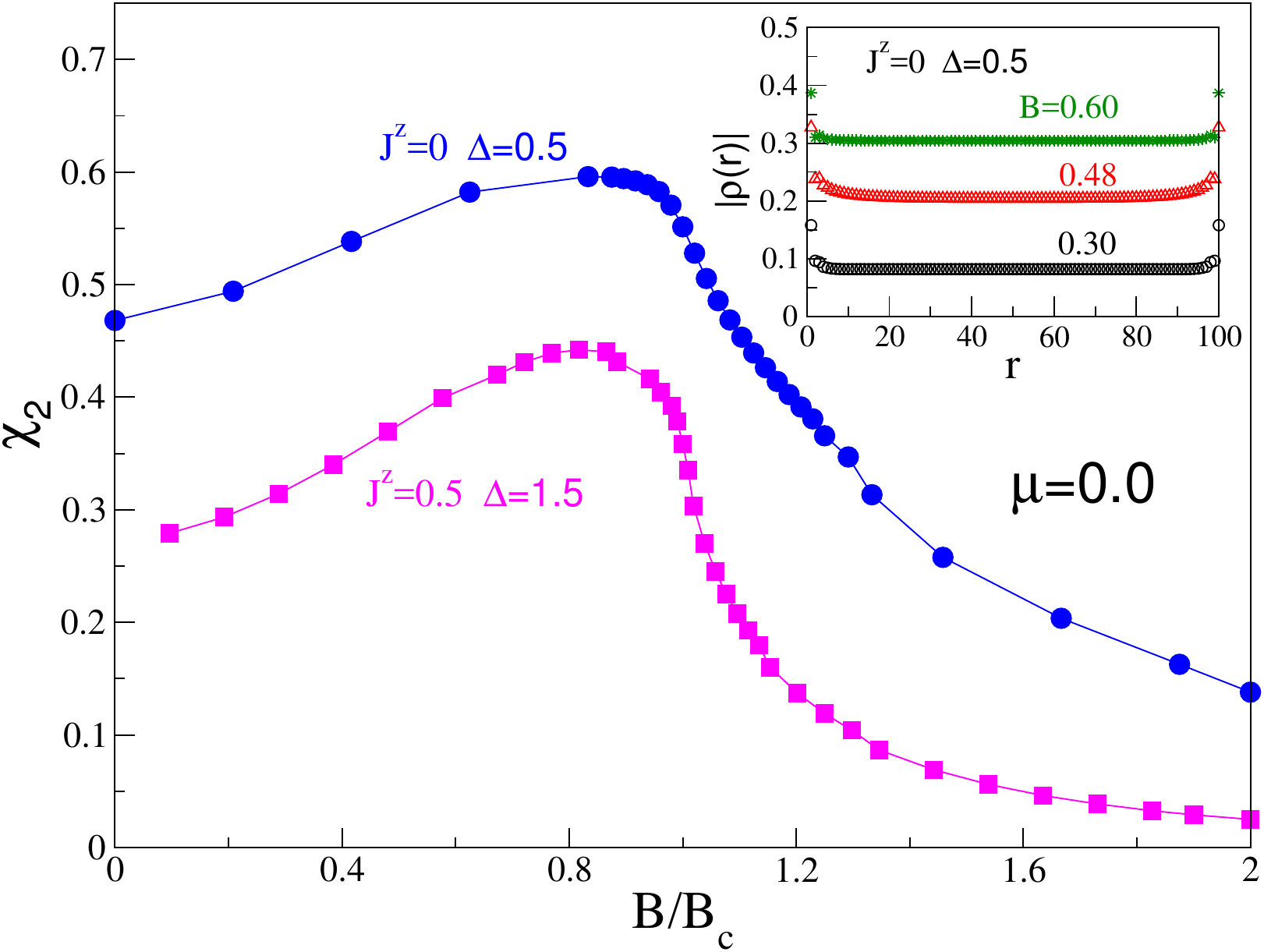}
\caption{Local staggered magnetic susceptibility (in~\eref{eq:sus})
of the second site nearest neighbor to the edge $\chi_2$ vs. $B/B_c$ both for
($\Delta = 0.5$, $\mu  = J^z = 0$) and ($\Delta = 1.5$, $J^z = 0.5$, 
$ \mu =0 $). In the inset, the spin density $\rho(r)$ for the 
whole system for $B = 0.3$, $0.48$ and $0.6$ with 
$\Delta = 0.5$, $\mu  = J^z = 0$ is shown.}
\label{fig4}
\end{center}
\end{figure}

  The Majorana modes are confined to the edge of the system, 
therefore we focus on the spin density of edge sites $i = 1$ 
and $2$ for $ \Delta=0.5 $, $ \mu=0 $ and $ J^z=0 $.  The 
spin density and  the local staggered  magnetic susceptibility 
is defined as
\begin{eqnarray}
\rho_i=2\left\langle S^z_i \right\rangle,  \\
\chi_i=\left|\frac{d\rho_i}{dB}\right|,
\label{eq:sus}
\end{eqnarray}
where, $ \left\langle S^z_i \right\rangle  $ are the longitudinal 
spin density at site $i$. The magnitude of spin density 
$ \rho_i $ for site $i = 1, 2$ increases at both the sites with $B$, 
it is positive at site $1$ and negative at site $2$. The magnitude 
of $ \rho_i $ at site $2$ is lower than that is at site $1$. However, 
both of them saturate with high staggered field, but $ \rho_1 $ 
continuously increases and there is no maxima for this function.  
The $ \chi_2 $ as a function of $B/B_c$ are plotted in the main~\fref{fig4} 
for ($J^z=0$, $\Delta=0.5$) and ($J^z=0.5$, $\Delta=1.5$) 
and both for $ \mu=0 $.  These two functions show a maxima near the 
transition. The $ \rho_i $ for the whole system for $J^z=0$, 
$\Delta=0.5$ and $\mu=0$ is shown in the inset of~\fref{fig4}. 
We notice that the variation of $ \rho_i $ is confined to the edge 
and the first few neighboring sites, and has constant value throughout 
the rest of system. The $ \rho_i $ for $J^z=0.5$, $\Delta=1.5$ and 
$\mu=0$ behaves in a similar manner.
%
\subsection{\label{Sec:E}Quadrupolar order parameter $P(r=1)$}
\begin{figure}
\begin{center}
\includegraphics[width=3.3 in]{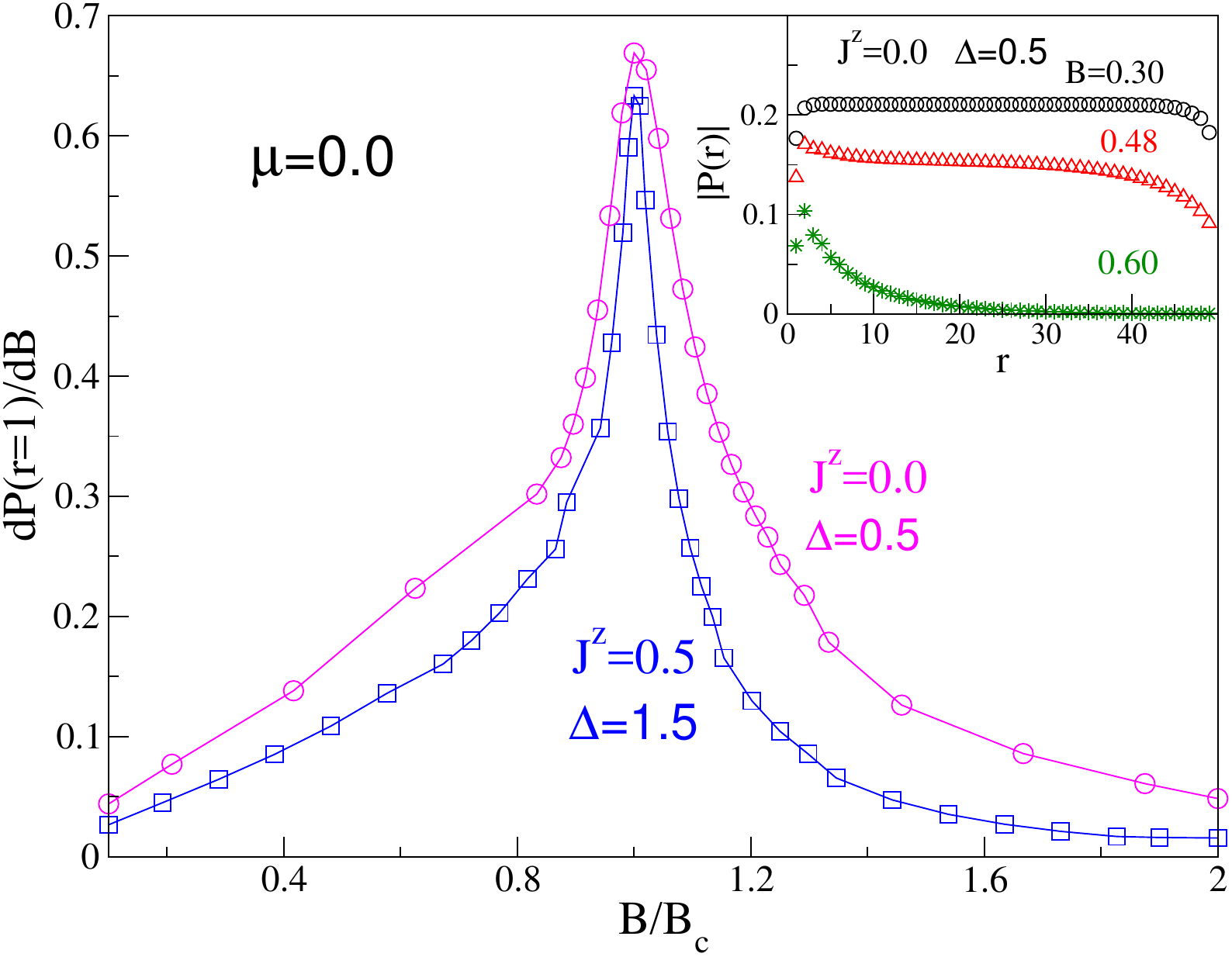}
\caption{ $ \frac{dP(r=1)}{dB} $ as a function of $\frac{B}{B_c}$ is shown
in the main figure for ($\Delta = 0.5$, $\mu  = J^z = 0$) and ($\Delta = 1.5$, 
$J^z = 0.5$, $ \mu =0 $). $P(r)$ (in~\eref{eq:pr}) is plotted for the whole system in 
the inset for $B = 0.3$, $0.48$ and $0.6$ with $\Delta = 0.5$, $\mu  = J^z = 0$.}
\label{fig5}
\end{center}
\end{figure}

The third term of the Hamiltonian in~\eref{eq:spin_ham} induces 
the spin quadrupolar/spin-nematic order in the system. In this 
phase $ \left\langle S^+\right\rangle  $ vanishes, whereas
$\left\langle S^+S^+\right\rangle $ has non-zero value, and two magnon 
pair formation is favored similar to superconductor system where 
two electrons form Cooper pair. The spin distance dependent 
quadrupolar order parameter is defined as
\begin{eqnarray}
P(r)&=& \left\langle S^x_iS^x_{i+r} \right\rangle -\left\langle 
S^y_iS^y_{i+r} \right\rangle, \nonumber \\
&=& \left\langle S^+_iS^+_{i+r} + S^-_iS^-_{i+r} \right\rangle.
\label{eq:pr}
\end{eqnarray}
$P( r )$ is the difference in the $X$ and $Y$ component of correlation 
$ \left\langle \vec{S}_i \cdot \vec{S}_{i+r} \right\rangle $. 
For $r = 1$, this quantity is very similar to the spin quadrupolar 
or superconducting order parameter of the spinless fermion model. 
The $P(r = 1)$ is calculated as a function of $B$ for 
($ \Delta = 0.5$, $J^z=0$) and ($\Delta = 1.5$, $J^z=0.5$) and both 
for $ \mu=0 $, and we notice that when $ \Delta $ dominates over $B$, 
the system goes from non-degenerate Ising phase to doubly degenerate 
states which favors Majorana edge state and the bulk phase goes to 
spin quadrupolar phase as $P(r) \neq 0$. The $P(r = 1)$ first 
increases with $B$, and then it saturates with higher $B$. The 
derivative of $P(r = 1)$ with $B$ for ($\Delta = 0.5$, $J^z=0$, 
$\mu=0$) and ($\Delta = 1.5$, $J^z=0.5$, $\mu=0$)  shows maxima at 
$B= 0.48$ and $B=1.04$ respectively as shown in the main~\fref{fig5}. 
The distance dependence of $P(r)$ as a function $r$ 
are shown for both the regime of the Majorana modes and the Ising 
states in inset of~\fref{fig5}. The reference site $i$ is the 
edge site of the chain. In the Majorana state P(r) shows long range 
behavior, however, it decays exponentially in case of Ising phase. 
%
\subsection{\label{Sec:F}Longitudinal structure factor $S(k)$}
\begin{figure}
\begin{center}
\includegraphics[width=3.3 in]{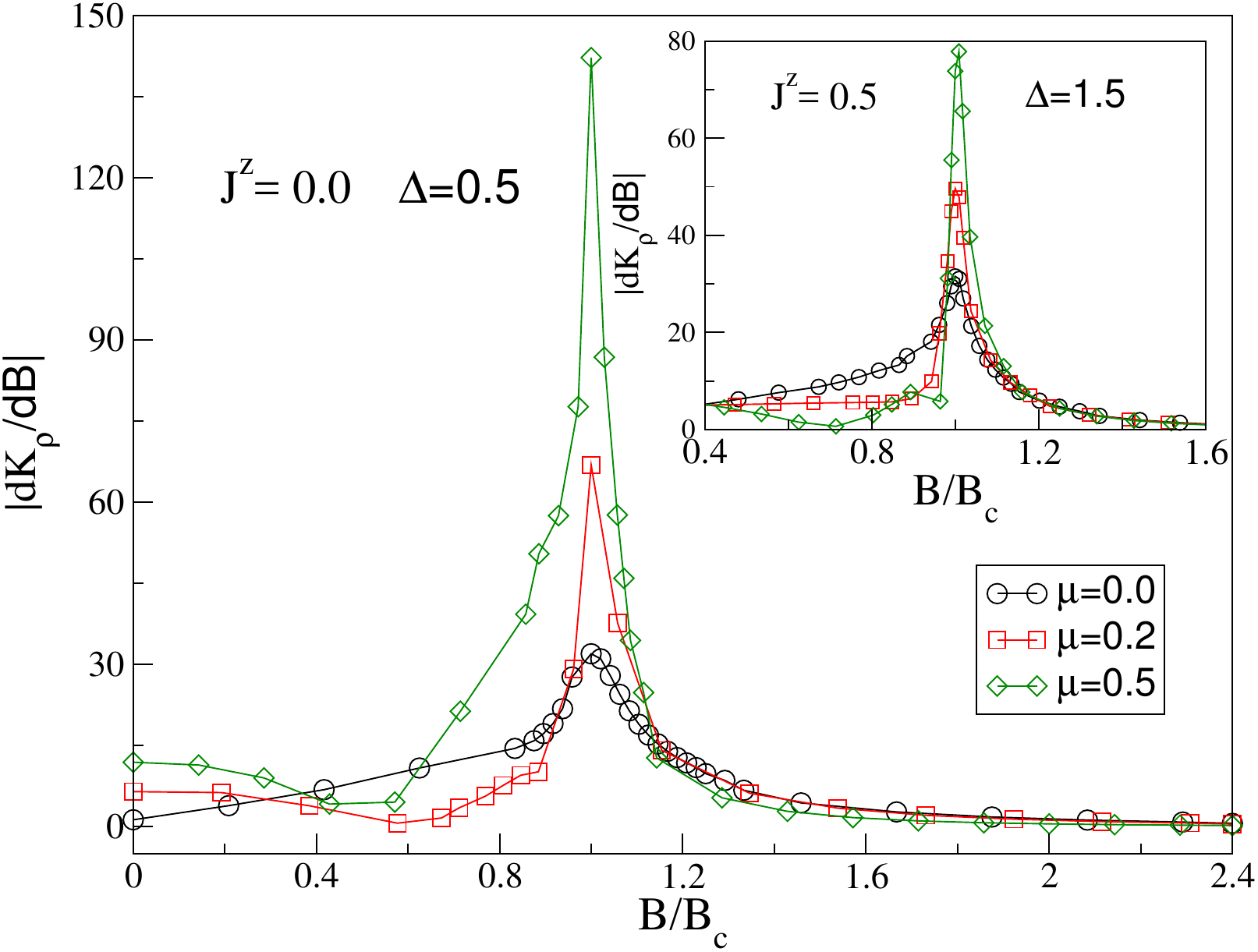}
\caption{Derivative of $K_\rho$ (in~\eref{eq:krho}) with the 
staggered magnetic field $B$, as a function of $ \frac{B}{B_c}$ 
is shown in the main figure for $\Delta = 0.5$, $J^z = 0$, and the 
inset shows the same for $\Delta = 1.5$, $J^z = 0.5$ for $\mu=0$, $0.2$ 
and $0.5$.}
\label{fig6}
\end{center}
\end{figure}

The longitudinal structure factor $S(k)$ is the Fourier transformation 
of $C(r)$ given in~\eref{spin_spin_corl}, and can be defined as
\begin{equation}
S(k)=\sum_r \left( \left\langle S^z_iS^z_{i+r}\right\rangle - 
\left\langle S^z_i\right\rangle \left\langle  S^z_{i+r}\right\rangle\right) e^{ikr}.
\end{equation}
Now let us define a quantity $ K_\rho $ in small $k$ limit defined as 
\begin{equation}
K_\rho=\frac{S(k)}{k/\pi};k\rightarrow 0,
\label{eq:krho}
\end{equation}
where $ k=\frac{2\pi m}{N}; $ $ m=0,\pm 1,\pm 2,\ldots,\pm \frac{N}{2}$. 
$ K_\rho $ is proportional to the Luttinger Liquid 
parameter~\cite{luttinger1,luttinger2,luttinger3,luttinger4}. We take 
the value of the function $ \frac{\pi S(k)}{k} $ at 
$ k=\frac{2\pi}{N} \, \left( m=1\right)  $. We calculate $ K_\rho $ for 
the ($ \Delta =0.5$, $J^z = 0 $) and ($ \Delta =1.5$, $J^z = 0.5 $) as 
a function of  $ \mu $ and $B$. The derivative of $ K_\rho $ as a 
function of $B$ shows maxima at  $B=0.48$, $0.52$, and $0.7$ for 
$\mu=0$, $0.2$ and $0.5$ at $J^z=0$ and $ \Delta=0.5 $. The derivative 
of $ K_\rho $ with $B/B_c$ for ($\Delta =0.5$, $J^z = 0$) and 
($\Delta =1.5$, $J^z = 0.5$) is shown for three different values of 
$\mu$ in the main and in the inset of~\fref{fig6}. The maxima of 
$\frac{dK_\rho}{dB} $ indicates the boundary between the Majorana and 
the Ising state. The extrapolated value of the transition point is very 
close to the transition point calculated from other criteria. It also 
shows that the critical value of $B$ for the transition calculated from 
$ \frac{dK_\rho}{dB} $ at a fixed $\Delta  $ increases with increasing $ \mu $. 
%
\subsection{\label{Sec:G}Entanglement spectrum in Ising and Majorana state}
As the Majorana mode is topological in nature, there is no 
well-defined order parameter. Therefore, direct measurement of 
this property is not possible~\cite{entanglement1,entanglement2,entanglement3}. 
Therefore, we study the ES of the reduced density matrix of the 
system to indirectly study the topological aspect. 
In this phase, all the states are either 
doubly or multiply degenerate~\cite{entanglement1, entanglement2}. 
The reduced density matrix of a system (half of the full system) can
be constructed in the GS of the full system by integrating out the 
environment degrees of freedom. The eigenvalues of the reduced 
density matrix are represented as $\lambda_i$. The Schmidt gap is defined as
$\Delta_S = \lambda_0 - \lambda_1$, where $\lambda_0$ and $\lambda_1$
are the largest and second largest eigenvalues of the reduced density
matrix. Topological phases are also characterized by $\Delta_S = 0$,
whereas it is finite in the trivial phase~\cite{entanglement2}. 
In a non-topological state the largest eigenvalue is non-degenerate 
and has mixed degenerate and non-degenerate eigenvalues~\cite{entanglement1, entanglement2}.

\begin{figure}
\begin{center}
\includegraphics[width=3.3 in]{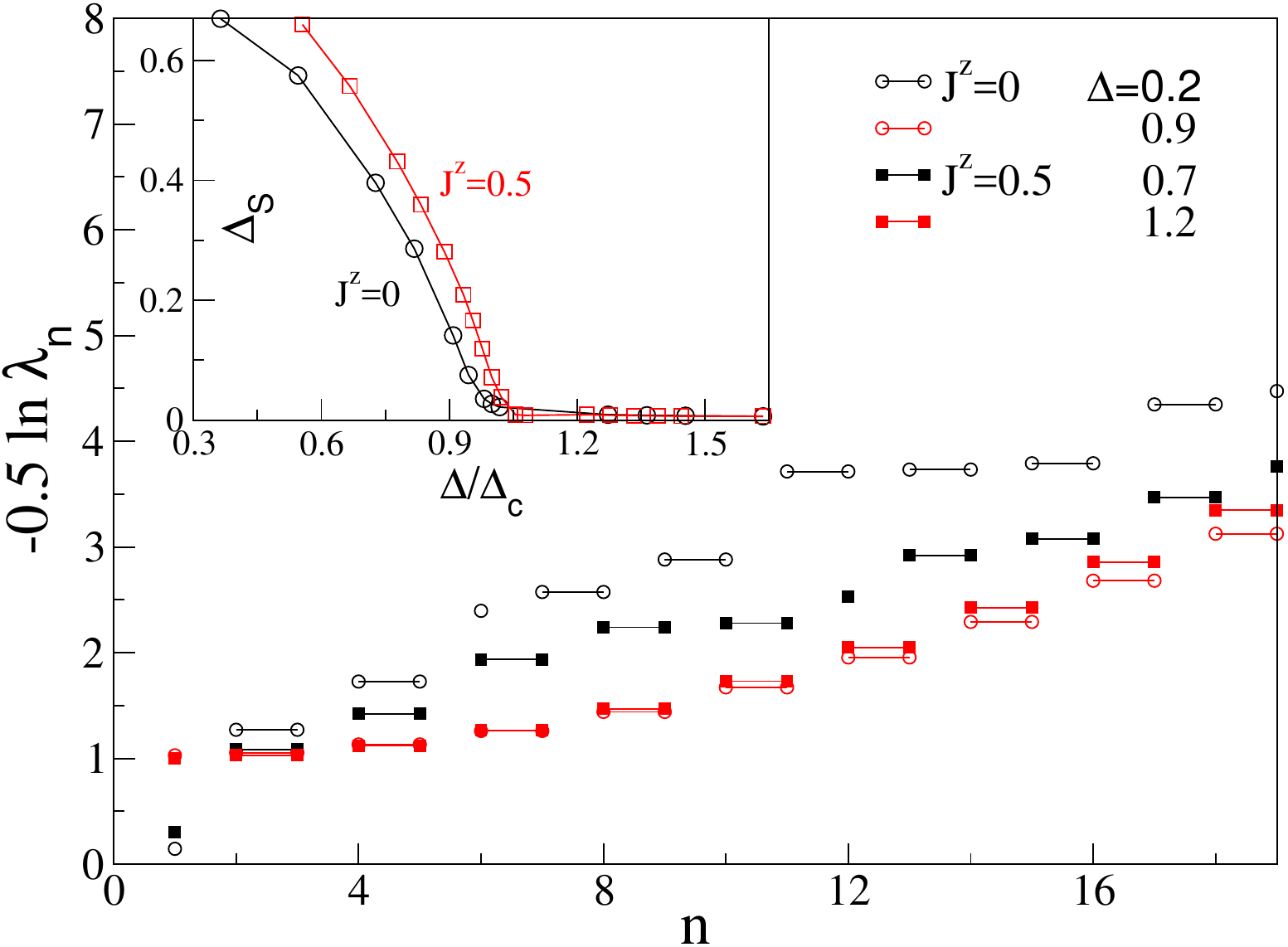}
\caption{The entanglement spectrum $\lambda_n$ is plotted for 
$\mu=0$ and $B=0.5$ in both the Ising and Majorana regime for 
$J^z =0$ and $J^z =0.5$ in the main figure. In the inset, the Schmidt gap
$ \Delta_S $ is shown as a function of $ \frac{\Delta}{\Delta_c}  $ for  
$B=0.5$, $ \mu=0 $ both for $J^z =0$ and $J^z =0.5$.}
\label{fig7}
\end{center}
\end{figure}

 The ES of the reduced density matrix is analyzed for a chain of $N=96$ spins with PBC 
in the deep Ising state, and in the deep Majorana state  for $J^z=0$ and $0.5$. We plot the 
Schmidt gap  as a function of $ \frac{\Delta}{\Delta_c} $ for 
$J^z =0$ and $J^z=0.5$ shown in the inset of~\fref{fig7} for 
$\mu=0 $ and $ B=0.5 $.  We notice that $\lambda_{n=0}$ 
is non-degenerate in the Ising state and Schmidt gap is finite 
in this regime but goes to zero in the Majorana state as shown in 
the inset of~\fref{fig7}. In the Ising phase  many of $\lambda_n$ 
are non-degenerate for  ($ J^z =0$,  $\mu=0$, $B=0.5$, 
$\Delta = 0.2 $) and ($J^z =0.5 $,  $\mu=0.0$, $B=0.5$, $\Delta = 0.7$). 
In this phase the ES is of mixed type. The spectrum 
is shown as open and filled symbols in the main~\fref{fig7} for 
$J^z=0$ and $0.5$ respectively. In the Majorana phase the $\lambda_{n=0}$ 
is triply degenerate, and for other higher $n$, these are either doubly 
or multiply degenerate as shown in the main~\fref{fig7} for 
($J^z =0$, $ \mu=0$, $B=0.5$, $\Delta = 0.9$) and ($J^z =0.5$,  $\mu=0$, 
$B=0.5$, $\Delta = 1.2$). The phase boundary of the system can be 
characterized at the point where Schmidt gap goes to zero and the whole 
spectrum becomes doubly or multiply degenerate. In the Majorana  phase 
first few largest $\lambda_n$ are independent of parameters.
\subsection{\label{Sec:H}MI phase boundary}
\begin{figure}
\begin{center}
\includegraphics[width=3.3 in]{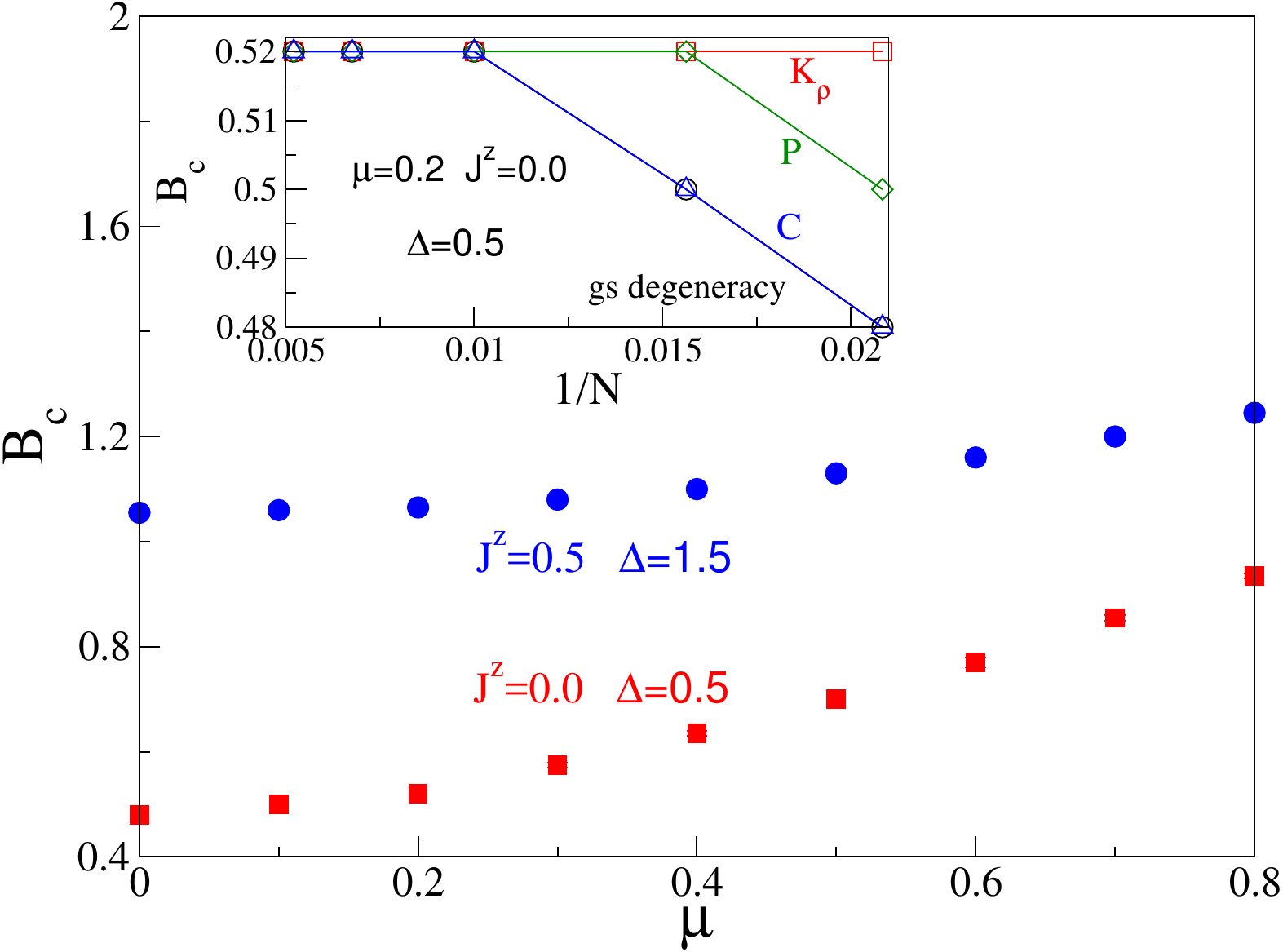}
\caption{The effect of $\mu$ on MI transition in the thermodynamic 
limit both for $J^z=0$, $ \Delta=0.5 $ and $J^z=0.5$, $ \Delta=1.5 $ 
are shown in the main figure. The finite size scaling of $B_c$ from 
different criteria for $J^z =0$, $\Delta=0.5$ and $ \mu=0.2 $ is 
shown in the inset.}
\label{fig8}
\end{center}
\end{figure}

In the last seven subsections, different criteria give us different 
MI phase boundary in a finite system size and the phase boundary 
$B_c$, calculated from different criteria have different finite size 
dependence. For $\mu=0.2$, the finite size scaling of the $B_c$ for 
four different criteria, i.e., ground state degeneracy, 
$ K_\rho $, $P(r=1)$ and $C(r=1)$ are shown in the inset of~\fref{fig8}. 
The ground state degeneracy and $C(r=1)$ have same finite size effect.
 We notice that the extrapolation from all the criteria leads to same 
$B_c=0.52$ in the thermodynamic limit. The effect of $ \mu $ on the MI 
transition point $B_c$  is shown in the thermodynamic limit for 
($J^z=0$, $ \Delta=0.5 $) and ($J^z=0.5$, $ \Delta=1.5 $) in the main~\fref{fig8}.
%
%
\section{\label{sec3}Discussion}
We have studied the helical liquid system and mapped this 
model into a $ XYZ $ spin-$1/2$ chain model. The Majorana-Ising 
transition is characterized by calculating the lowest 
excitation gap of the model Hamiltonian in~\eref{eq:spin_ham} 
on a chain geometry. In the Majorana state, the system has 
finite gap for a finite system which decays exponentially with 
the system size $N$. The closing of the gap in the thermodynamic 
limit is consistent with the study by Sela et al.~\cite{sela}. 
Our aim of this paper is to explore various criteria to 
characterize Majorana mode other than closing of the lowest excitation 
gap, and the accurate determination of the MI transition boundary. 
We have calculated various quantities, e.g., $\rho_i$, $C( r )$, $P( r )$, 
$K_\rho$,  $\Delta_S$.  We have shown that the phase boundary 
calculated from the various criteria  are the same for given value
 of $\Delta$ and $\mu$ in the thermodynamic limit. 
We have shown that in strong repulsive interaction limit $J^z>0 $, 
Majorana mode occurs at higher value of $ \Delta $ than the 
non-interacting case $ \left( J^z=0\right)  $, which is 
consistent with the study of  Gangadharaiah et al.~\cite{suhas}, 
where they have shown that the repulsive interaction weakens the 
Majorana modes. The $J^z$ term  of~\eref{eq:spin_ham} is 
similar to the  repulsive interaction term of the spinless fermion 
model in~\eref{eq:fermion_ham}. In the mean field limit, $J^z$ 
term reduces to effective $ \mu $. We have noticed that the 
Majorana state occurs for 
$(\mu-B)^2 < \Delta^2$, therefore, any change in $\mu$ changes the 
value of critical field $B_c$. The $P(r)$ is long range in the 
Majorana state, whereas it decays exponentially in the 
Ising phase as shown in the inset of~\fref{fig5}. In the Majorana 
state, the bulk of system shows quadrupolar/or spin nematic phase 
like behavior. The local magnetic susceptibility $ \chi $ of the 
edge sites shows maxima near the phase boundary.

We have also studied the ES of this model, and it is shown that the 
Ising phase has finite Schmidt gap $ \Delta_S $, and non-degenerate 
eigenvalues are present in the spectrum of the reduced density matrix 
of the system, whereas the topological aspect of Majorana state is 
characterized by the doubly degenerate eigenvalues and zero Schmidt 
gap~\cite{entanglement1,entanglement2}. We have also shown that ES 
of the reduced density matrix of the ground state shows  double and 
multiple degeneracy  in the Majorana state. We have also noticed 
threefold degeneracy in the largest eigenvalues in this state in the 
thermodynamic limit. Degeneracy of all the eigenvalues is very similar 
to the study by Pollmann et al.~\cite{entanglement1} to distinguish 
the topological and trivial phase of $ S=1$ system. The first few 
eigenvalues of ES in the Majorana state is almost independent of 
parameters as shown in~\fref{fig7}. 

In conclusion, we have studied the helical liquid phase in one 
dimensional system. This system shows the Majorana-Ising transition, 
and the phase boundary is calculated using various criteria. The 
topological aspect of the Majorana modes is studied for helical model, 
and the closing of the Schmidt gap and degeneracy of full spectrum of 
reduced density matrix can also be used to characterize the phase 
boundary of Majorana-Ising transition.  This model is one of the most 
interesting and a general model for spin-$1/2$ systems.  Our 
study shows that an anisotropic spin-$1/2$ chain can be a good 
candidate to observe the Majorana modes and MI transition. The local 
experimental probe like neutron magnetic resonance can be used to 
measure the local spin density at the edge of the sample. 
\ack
MK thanks Z. G. Soos and Sumanta Tewari for their valuable comments.
MK thanks DST for a Ramanujan Fellowship SR/S2/RJN-69/2012 and funding 
computation facility through SNB/MK/14-15/137. SS thanks the DST 
(SERB, SR/S2/LOP-07/2012) fund  and SNBNCBC for supporting the visit.
 
\section*{References}
\bibliography{ref_MI}
\end{document}